\documentclass[sigconf]{acmart}
\usepackage{xcolor}         % colors
\usepackage{multirow}
\usepackage{multicol}
\usepackage{colortbl}
\usepackage{adjustbox}
\usepackage{caption}
\usepackage{subcaption}
\usepackage{amsmath}
\usepackage{todonotes}
\usepackage{pifont}
\usepackage{array}
\usepackage{ragged2e}
\usepackage{booktabs}
\usepackage{wrapfig}
\usepackage{changepage,threeparttable} % for wide tables
\usepackage{enumitem} % for adjusting itemize indent
\definecolor{mygreen}{HTML}{097969}
\newcommand{\method}{CovDocker}
\usepackage{balance} 

\usepackage[arxivVer]{optional}
\newcommand\arxivurl{https://github.com/PoloWitty/CovDocker/blob/4170fd2728b7bccd16c00a52e38e13dd58dd5db6/KDD25__CovDocker__appendix.pdf}
\AtBeginDocument{%
  }

\copyrightyear{2025}
\acmYear{2025}
\setcopyright{acmlicensed}
\acmConference[KDD '25]{Proceedings of the 31st ACM
SIGKDD Conference on Knowledge Discovery and Data Mining V.2}{August 3--7,
2025}{Toronto, ON, Canada}
\acmBooktitle{Proceedings of the 31st ACM SIGKDD Conference on Knowledge
Discovery and Data Mining V.2 (KDD '25), August 3--7, 2025, Toronto, ON, Canada}
\acmDOI{10.1145/3711896.3736896}
\acmISBN{979-8-4007-1454-2/2025/08}

\begin{document}

\title{CovDocker: Benchmarking Covalent Drug Design with Tasks, Datasets, and Solutions}


\newcommand{\hustAffiliation}{
    \affiliation{%
      \institution{Huazhong University of Science and Technology}
      \city{Wuhan}
      \state{Hubei}
      \country{China}
    }
}

\newcommand{\msraAffiliation}{
    \affiliation{%
      \institution{Microsoft Research AI for Science}
      \city{Beijing}
      % \state{Beijing Shi}
      \country{China}
    }
}

\author{Yangzhe Peng}
\authornote{Both authors contributed equally to this research.}
\email{yangzhepeng@hust.edu.cn}
\orcid{0009-0006-2873-1659}
\author{Kaiyuan Gao}
\authornotemark[1]
\email{im_kai@hust.edu.cn}
\orcid{0009-0002-8862-8320}
\hustAffiliation

\author{Liang He}
\email{lihe@microsoft.com}
\orcid{0000-0002-6394-8531}
\msraAffiliation

\author{Yuheng Cong}
\email{alexcong@hust.edu.cn}
\orcid{0009-0006-3833-8850}
\hustAffiliation

\author{Haiguang Liu}
\email{haiguang.liu@microsoft.com}
\orcid{0000-0001-7324-6632}
\msraAffiliation

\author{Kun He}
\authornote{Corresponding authors}
\email{brooklet60@hust.edu.cn}
\orcid{0000-0001-7627-4604}
\hustAffiliation

\author{Lijun Wu}
\authornotemark[2]
\email{lijun\_wu@outlook.com}
\orcid{0000-0002-3530-590X}
\affiliation{%
  \institution{Shanghai Artificial Intelligence Laboratory}
  \city{Shanghai}
  \country{China}
}

\renewcommand{\shortauthors}{Yangzhe Peng et al.}

\begin{abstract}
Molecular docking plays a crucial role in predicting the binding mode of ligands to target proteins, and covalent interactions, which involve the formation of a covalent bond between the ligand and the target, are particularly valuable due to their strong, enduring binding nature. 
However, most existing docking methods and deep learning approaches hardly account for the formation of covalent bonds and the associated structural changes.
To address this gap, we introduce a comprehensive benchmark for covalent docking, \method, 
which is designed to better capture the complexities of covalent binding. 
We decompose the covalent docking process into three main tasks: reactive location prediction, covalent reaction prediction, and covalent docking. By adapting state-of-the-art models, such as Uni-Mol and Chemformer, we establish baseline performances and demonstrate the effectiveness of the benchmark in accurately predicting interaction sites and modeling the molecular transformations involved in covalent binding. These results confirm the role of the benchmark as a rigorous framework for advancing research in covalent drug design.
 It underscores the potential of data-driven approaches to accelerate the discovery of selective covalent inhibitors and addresses critical challenges in therapeutic development.

\end{abstract}

% TLDR: This paper introduces a novel benchmark dataset created by refining two existing covalent binding datasets for deep learning applications. Three new tasks are proposed: reaction site prediction, reaction prediction, and cov-docking, which advance traditional models and tasks in the field. We utilize the Uni-Mol and Chemformer model, achieving better performances than traditional method. These contributions set new benchmarks and encourage further developments in covalent drug design, aiming to accelerate advancements in the field.
%We present new benchmark datasets for deep learning in covalent drug design, introducing three advanced tasks and outperforming traditional methods with Uni-Mol and Chemformer models, setting new benchmarks for further advancements in the field.





\maketitle

\newcommand\kddavailabilityurl{https://doi.org/10.5281/zenodo.12805810}

\ifdefempty{\kddavailabilityurl}{}{
\begingroup\small\noindent\raggedright\textbf{KDD Availability Link:}\\
% please change the following context to include multiple artifacts if necessary.
The preprocessed dataset of this paper has been made publicly available at \url{\kddavailabilityurl}.
The source code of this paper is available at \opt{arxivVer}{\url{https://github.com/PoloWitty/CovDocker}}\opt{kddCRVer}{\url{https://doi.org/10.5281/zenodo.15526093}}.
\endgroup
}

%%%%%%%%%%%%%%%%%%%%%%%%%%%%%%%%%%%%%%%%%%%%%%%%%%%%%%%%%%%%%%%%%%%%%%%%%%%%%%%%%%%%%
%%%%%%%%%%%%%%%%%%%%%%%%%%%%%%%%%%%%%%%%%%%%%%%%%%%%%%%%%%%%%%%%%%%%%%%%%%%%%%%%%%%%%

\section{Introduction}

\begin{figure}
% \vspace{-1cm}
\centering
\includegraphics[width=\linewidth]{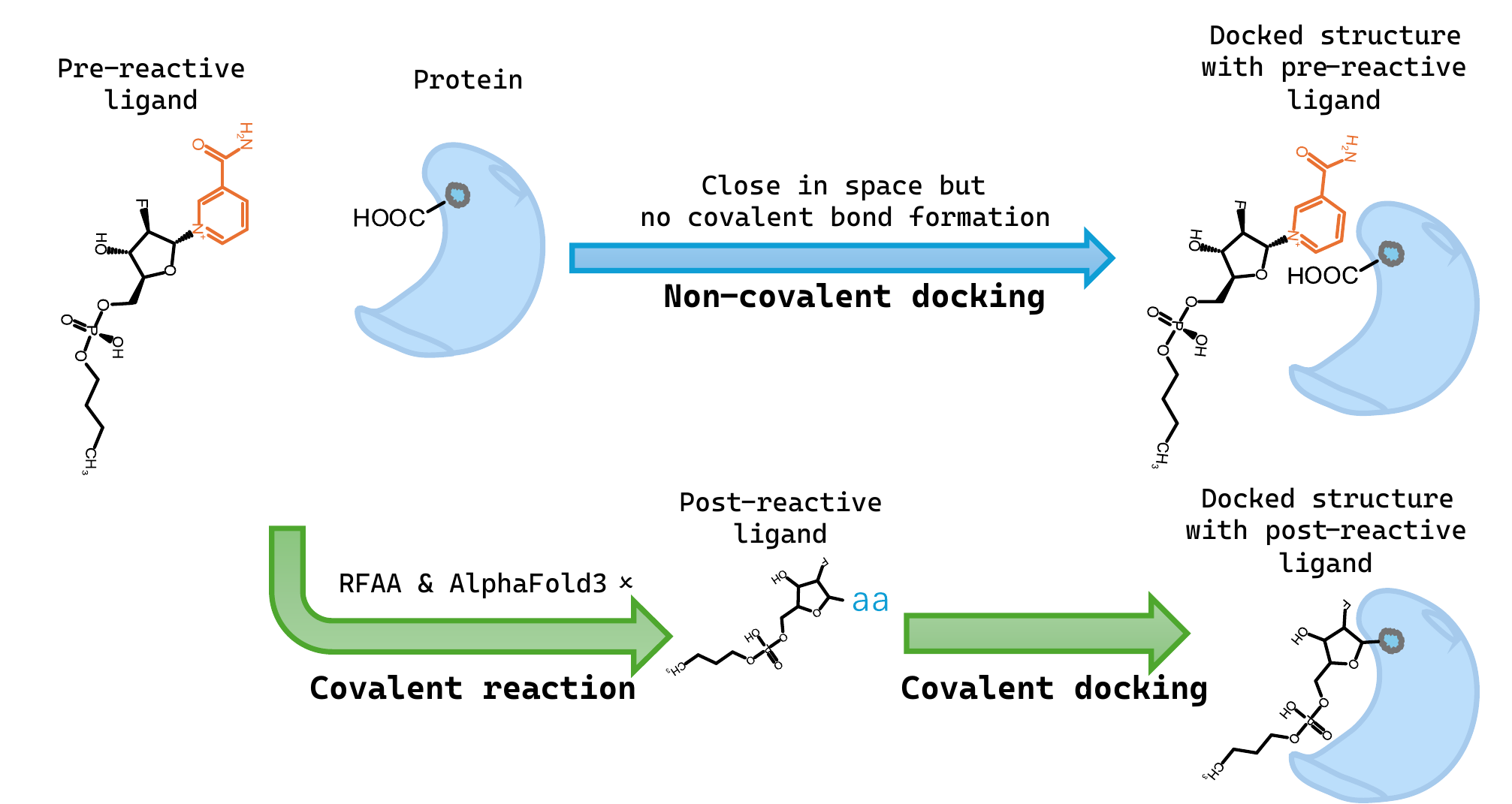}
\caption{Graphical overview of different methodological approaches in covalent drug design. 
Traditional non-covalent docking methods, which are well-established in molecular docking, can only predict ligand-protein interactions where the ligand and protein are spatially close but do not account for the formation of a covalent bond. 
In contrast, methods such as RFAA and AlphaFold3 can model covalent docking; however, they overlook the critical covalent reaction step, which is essential for accurate covalent drug design.}
\label{fig:motivation-fig}
% \vspace{-1em}
\end{figure}

Molecular docking, a computational technique used to predict 
the binding mode of ligands to target proteins, plays an increasingly important role in the drug discovery process ~\cite{morris1996distributed,morris2008molecular,fan2019progress}. Among the various docking scenarios, covalent protein–ligand docking is notable for forming a covalent bond between the ligand and the target protein, resulting in strong and durable interactions~\cite{ai-driven-SARS2-covalent-drug-design,london2014,zhu2014docking}. In recent years, covalent inhibitors have garnered significant attention for their potential to improve potency, selectivity, and prolonged target engagement compared to non-covalent counterparts~\cite{baillie2016targeted,singh2011resurgence,zhu2014docking}. Furthermore, it is estimated that up to 30\% of clinically approved drugs act through a covalent bonding mechanism~\cite{gao2022}, underscoring the clinical relevance of covalent drug design.

However, traditional methods in this field struggle to effectively model covalent binding and the associated structural changes effectively. 
Key limitations include:
(1) They tend to be computationally intensive, often relying on the energy function optimization~\cite{ouyang2013}. 
(2) They require substantial manual expertise to design energy functions tailored to specific reactions~\cite{wu2022a} or to employ manual alignment strategies~\cite{bianco2016}. 
(3) Many focus only on reactive amino acids without considering the entire binding pocket~\cite{bianco2016},
(4) The field is hindered by a lack of open and reproducible resources, with many methods being proprietary or lacking accessible code. 

\begin{figure*}[ht]
    \centering
    \includegraphics[width=\textwidth]{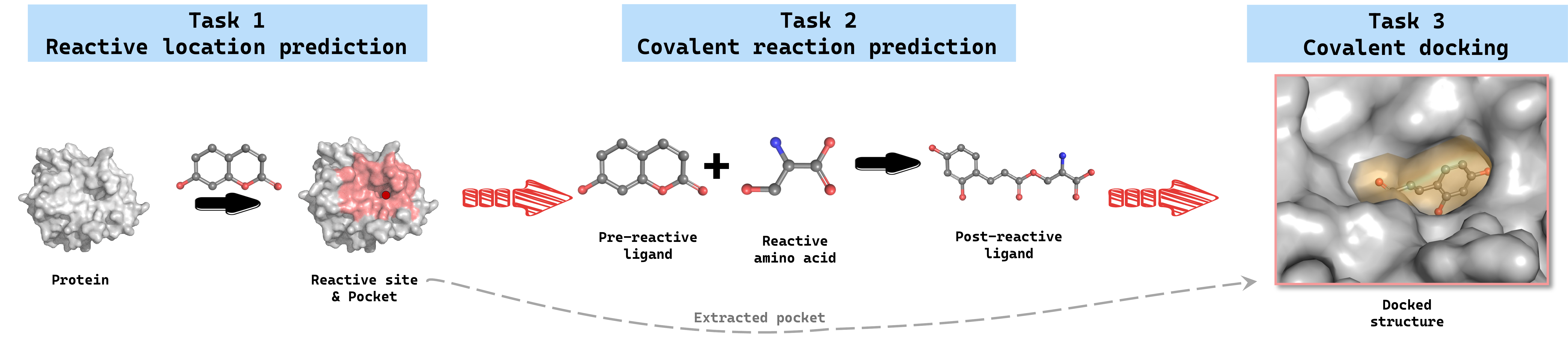}
    \caption{Illustration of the three proposed tasks in our covalent docking framework. 
    \textit{Task 1 - Reactive location prediction}: Given the holo protein structure and a pre-reactive ligand, the goal is to identify both the pocket and the correct reactive residue site. 
    \textit{Task 2 - Covalent reaction prediction}: For a pre-reactive ligand and a reactive amino acid, the task is to predict the post-reactive ligand after the covalent bond formation. 
    \textit{Task 3 - Covalent docking}: Given the apo structure of the post-reactive ligand, the goal is to predict the correct ligand pose after covalent docking.}
    \label{fig:task}
    % \vspace{-1em}
\end{figure*}

Recent advances in deep learning have shown promise in various molecular modeling tasks~\cite{krishna2023,zhou2023}, yet their application to covalent docking remains limited. Many existing deep learning methods for docking primarily focus on non-covalent interactions and fail to model the reactive events and structural transformations intrinsic to covalent binding~\cite{corso2023, stark2022, pei2024fabind, gao2024fabind+, jiang2025posex}. As illustrated in Figure~\ref{fig:motivation-fig},  
when traditional non-covalent docking methods are applied to covalent drug design, they can predict docking structures where the ligand and protein are spatially close but fail to capture the formation of a covalent bond.  
Furthermore, while recent models like RFAA~\cite{krishna2023} and AlphaFold3~\cite{abramson2024} have made significant strides in modeling covalent bonds, they overlook critical steps in the covalent reaction process. This makes them incapable of handling issues such as structural changes caused by leaving groups. In addition, the lack of specialized datasets and tailored tasks for covalent docking further hinders the development of accurate and robust deep learning models in this field.

To address these challenges, we introduce \textbf{\method}, a comprehensive benchmark designed to advance covalent drug design by incorporating tailored tasks and datasets for deep learning.
We collects and integrates high-quality datasets focusing on covalent interactions, systematically organizing them into three key tasks.
We also introduce new evaluation metrics to rigorously assess the performance of various models. With \method, we offer a unified framework that integrates these tasks and sets the stage for future advancements in covalent docking using deep learning models.

As illustrated in Figure~\ref{fig:task}, we decompose the covalent drug design process into three main tasks: (1) \textit{Reactive location prediction}, which involves identifying both the pocket in the protein and the specific residue site most likely to undergo a covalent chemical reaction; (2) \textit{Covalent reaction prediction}, which determines the product of a covalent reaction by predicting atom connectivities and functional groups formed between the residue and the molecule; and (3) \textit{Covalent docking (cov-docking)}, which predicts the low-energy complex conformation of the covalently bound molecule within the extracted pocket.

To tackle these tasks, we leverage cutting-edge models like Uni-Mol and Chemformer. 
Specifically, we use Uni-Mol Block~\cite{zhou2023} for reactive location prediction, incorporating a cross-attention mechanism to accurately identify the potential pockets and target residues site. 
For covalent reaction prediction, we fine-tune Chemformer~\cite{irwin2022} as a sequence-based model to forecast covalent reaction outcomes. Finally, we adapt the Uni-Mol docking model to train a site-specific docking method with auxiliary covalent constraints to enhance binding accuracy. 
Our extensive empirical evaluation with \method shows promising results, setting robust baselines for future research in covalent drug design.

Our main contributions can be summarized as follows:
\begin{itemize}[leftmargin=2em, labelsep=0.5em, topsep=0.25em]
% Conceptually
\item We introduce \method, a novel conceptual framework designed to address the complex challenges in covalent protein-ligand docking, an under-explored area in computational drug discovery.  
% Technically
\item We establish a comprehensive benchmark for covalent drug design, consisting of modified tasks and adapted datasets for training and testing deep learning models. We break down the covalent docking process into three main tasks and establish new evaluation metrics.
% Empirically
\item Comprehensive experiments demonstrate the effectiveness of our technical innovations and highlight the transformative potential of deep learning in covalent drug design, paving the way for more potent and selective therapeutic agents.
\end{itemize}

%\vspace{-1em}
\section{Related Work}

\subsection{Molecular Docking}
Traditional molecular docking methods typically rely on computational algorithms to search/explore the conformation space of ligands and receptors, and predict their binding interactions~\cite{oyedele2023,elokely2013docking,ewing1997critical}. 
Some commonly employed traditional docking approaches include: Lamarckian Genetic Algorithm, used in AutoDock to explore binding configurations and evaluate affinities using a grid-based energy scoring function~\cite{morris1998automated, morris2008using, cosconati2010virtual}; and Dock, which employs a geometric matching algorithm followed by energy minimization to predict ligand binding poses~\cite{ewing2001dock}.
These approaches can be broadly categorized into two types: site-specific docking~\cite{zhang2023efficient,zhou2023}, which predicts ligand conformation within a given pocket, and blind docking~\cite{corso2023,lu2022tankbind,pei2024fabind,gao2024fabind+}, which docks the ligand onto the entire protein without prior knowledge of the pocket location. Our work decomposes blind docking into two steps: pocket identification and site-specific docking.

\subsection{Covalent Docking}
\textbf{Traditional Covalent Docking Methods.}  The development of covalent docking lags behind compared to the non-covalent docking, partly due to the complexity of modeling irreversible bonds essential for covalent drug design.
Traditional covalent docking approaches include CovalentDock~\cite{ouyang2013}, which integrates an empirical model for free energy estimation, and DOCKovalent~\cite{london2014}, which confines the ligand’s warhead near the reactive residue to reduce conformational space. DOCKTITE~\cite{scholz2015} creates a pseudo receptor side chain connection for precise interactions, while AutoDock4 (cov)~\cite{bianco2016} aligns the ligand with a rigid protein backbone. CDOCKER (cov)~\cite{wu2022a} enhances docking with a customizable covalent bond grid potential, optimizing free energy change for bond formation. Most recently, HCovDock\cite{wu2023} integrates a ligand sampling technique based on incremental construction and a scoring function that incorporates covalent bond-based energy.

% \paragraph{Reactive binding site prediction}  Existing methods typically evaluate the endpoint of reactions by modeling ligands in their bound state. Recently, Reactive Docking~\cite{bianco2023} has been introduced to not only predict the correct reactive residue but also identify the optimal reactive ligand, which mirrors our task on reactive site prediction. However, this method utilizes a straightforward approach, directly employing a non-covalent docking tool combined with flexible side chain docking that enumerates each potential individual residue. A ligand and residue are considered reactive if the distance between their reactive atoms in the lowest energy pose is less than 2.0 Å. Despite reporting favorable results on their evaluation set, the extensive enumeration of all possible individual residues may hinder its applicability in real-world virtual screening scenarios. A similar task is ligandable site prediction, which identifies all ligandable binding sites in a protein instead of predicting the reactive site for a given pre-reactive ligand. DeepCys~\cite{liu2024a} has been proposed to predict proteome-wide covalent ligandabilities directed at cysteines.

\noindent\textbf{Covalent Docking Datasets.}  Prior datasets primarily served as evaluation sets in covalent docking, and the data scale is very limited. 
For example, \citet{scarpino2018} features 207 Cys-bound complexes, renowned for its quality standards and is widespread used for traditional covalent docking tools. Similarly, BCDE set~\cite{wen2019} only contains 330 examples (245 Cys and 85 Ser residues).
Some efforts have focused on refining these datasets by combining, filtering, and diversifying the data to better evaluate the performance~\cite{moitessier2016, wei2022}. Several dedicated web resources for covalent complexes have been developed, such as the cBinderDB (no longer accessible) and the CovalentInDB~\cite{du2021}, which do not primarily focus on structural data. Recently, some larger screening databases have been built, such as CovPDB~\cite{gao2022} and CovBinderInPDB~\cite{guo2022}, which are based on structural data from the PDB, containing 2,261 and 3,555 complexes, respectively. Due to their larger size, the two sets have been used for some studies of ligand-able cysteine sites based on machine learning~\cite{liu2024a, gao2023}.

Existing covalent docking tools face two key limitations: (1) restricted code transparency (closed-source implementations or partial data release), and (2) dependence on evaluation-only datasets that lack preprocessed structural details and training-evaluation splits. 
While recent databases like CovPDB and CovBinderInPDB provide structural insights for inhibitor screening, their native formats remain incompatible with machine learning workflows. 
In contrast, \method{} introduces three critical improvements: (1) full reproducibility through open-source code, preprocessed datasets, and trained weights; (2) systematic conversion of structural databases into ML-ready benchmarks with dedicated training/evaluation sets and diverse reaction mechanisms; and (3) extended amino acid coverage beyond cysteine-centric systems. Further quantitative comparisons of \method{} with existing methods are provided in Section~\ref{sec: comparison}.

\section{\method: Covalent Drug Design Modeling}
In this section, we introduce the \method{} benchmark design, which defines three key tasks for covalent drug design, along with baseline solutions for each task.

\subsection{Method Overview}

For covalent docking, we consider two docking scenarios, site-specific docking and  blind docking. In site-specific docking, both the target pocket and reactive site are provided, and the goal is to predict the molecular docking pose. 
This scenario is the more commonly used setting in molecular docking studies. 
On the other hand, blind docking presents more challenges, as it only provides the full protein structure and a randomized ligand conformation, and the task is to predict the docking pose without prior knowledge of the binding site. Blind docking is particularly important when dealing with new or previously uncharacterized disease proteins.

Our approach follows the blind docking setting, where we divide the covalent docking problem into three clearly defined tasks for deep learning models: (1) reactive location prediction, (2) covalent reaction prediction, and (3) covalent docking pose prediction. 
The site-specific docking contains the last two tasks, as the pocket and reactive site are predefined. Figure~\ref{fig:task} provides an overview of these tasks and their relationships. 

\begin{figure*}
    \centering
    \includegraphics[width=\textwidth]{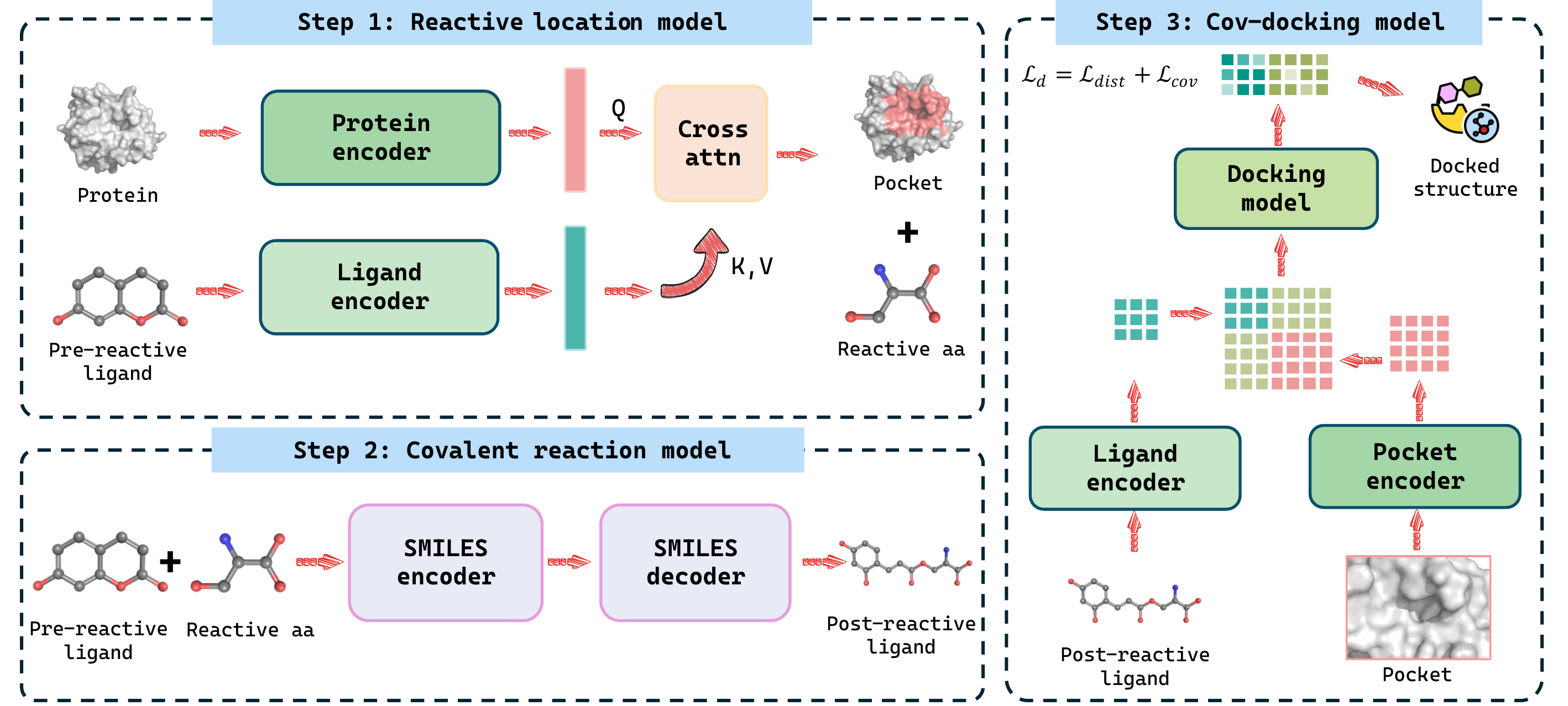}
    \caption{Illustration of our proposed models. All green blocks represent the Uni-Mol~\cite{zhou2023} blocks, with varying shades indicating different functionalities.}
    \label{fig:models}
    % \vspace{-0.3cm}
\end{figure*}

\subsection{Task 1: Reactive Location Prediction}
\label{sec:Task 1}
The first step in molecular docking is to predict the ligand-binding pocket on the protein. 
For covalent docking, this task is extended to predict not only the binding pocket but also the specific reactive site that will form a covalent bond with the ligand.

Formally, given a protein $\mathcal{P}$ with its holo structure $\mathcal{R^P}$, and the (pre-reactive) ligand $\mathcal{M}$ with a random conformation $\mathcal{R^M}$, the goal of reactive location prediction task is to predict both the pocket center $\mathcal{P}^c$ and the reactive site $\mathcal{P}^s$ using a prediction model $M_{l}$:
\begin{equation}
    (\mathcal{P}^c, \mathcal{P}^s) = M_l(\mathcal{R^P}, \mathcal{R^M}).
\end{equation}
Here, $\mathcal{P}^c$ denotes the pocket center, which is predicted as a regression task (the coordinate of the pocket center), and $\mathcal{P}^s$ represents the residue ID of the reactive site, predicted as a classification task. 
The pocket is defined as a circle with radius $20$\AA~ around the pocket center $\mathcal{P}^c$ during inference, and only the residues within this region are considered for covalent docking pose prediction in Task 3. 

During training, the ground-truth pocket is defined as the residues within a $10$\AA~radius of each ligand atom. 
The reactive site $\mathcal{P}^s$ (the residue that can have covalent interaction with ligand) will be used for Task 2, where we predict the covalent reaction. 
% Our definition of the pocket center is the geometric center of all pocket amino acid CA carbons in the protein, determined based on the 10 Å distance threshold from the ligand.

To model this task, we use Uni-Mol~\cite{zhou2023}, a deep learning framework for molecular representation learning. 
Uni-Mol includes separate protein and molecule encoders, and we modify the protein encoder to operate at the residue level (instead of atom level) for our task. 
As illustrated in Figure~\ref{fig:models}, we incorporate a cross-attention layer on top of the two encoders to integrate the protein and ligand representations. 
This is followed by two separate prediction heads to estimate the pocket center $\mathcal{P}^c$ and the reactive site $\mathcal{P}^s$, with the corresponding losses defined as $\mathcal{L}_{c}$ and $\mathcal{L}_{s}$. The final loss for Task 1 is then constructed by 
\begin{equation}
    \mathcal{L}_{p} = \mathcal{L}_{c} + \alpha \mathcal{L}_{s},
\end{equation}
where $\alpha$ is the loss weight. For further details on the feature construction, the modified Uni-Mol model, and loss functions, please refer to Appendix~\opt{arxivVer}{\ref{sec:app_models}}\opt{kddCRVer}{\href{\arxivurl}{B}}.

\subsection{Task 2: Covalent Reaction Prediction}
\label{sec:Task 2}
Once the reactive site (the residue) has been predicted, the next task is to model the covalent reaction that occurs when the ligand forms a covalent bond with the protein.  This task predicts the post-reactive ligand (the product of the covalent reaction).

As shown in Figure~\ref{fig:task}, given the pre-reactive ligand $\mathcal{M}$ and the reactive site $\mathcal{P}^s$, and the goal is to predict the post-reactive ligand (product) $\mathcal{M}'$ through a deep learning model $M_r$:
\begin{equation}
    \mathcal{M'} = M_r(\mathcal{M}, \mathcal{P}^s). 
\end{equation}
An important difference from the way in Task 1 is that the protein is represented in residue level, here we model the reactive site (residue) $\mathcal{P}^s$ in atom level, since we aim to predict all atoms of the generated post-reactive ligand. 

For this task, we represent all the molecules (pre-reactive ligand, residue, post-reactive ligand) in their sequence format using Simplified Molecular-Input Line-entry System (SMILES)~\cite{weininger1988smiles}. Then we use the Transformer-based model, Chemformer~\cite{irwin2022}, which is an encoder-decoder framework and has been demonstrated well in reaction prediction task, to model the covalent reaction prediction. The loss function used is a typical cross-entropy loss $\mathcal{L}_r$  for sequence generation.

A simplified workflow for this task is shown in Figure~\ref{fig:models}. Further modeling details and loss formulations are available in Appendix~\opt{arxivVer}{\ref{sec:app_models}}\opt{kddCRVer}{\href{\arxivurl}{B}}.

\begin{figure*}[t]
    \centering
    \includegraphics[width=\textwidth]{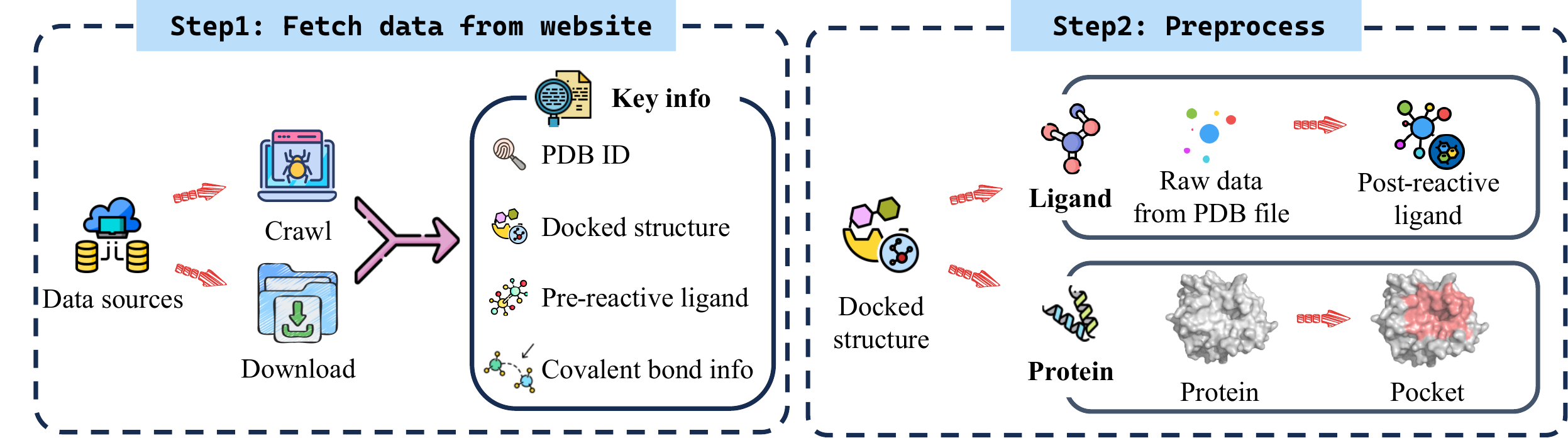}
    \caption{Overview on the data construction of \method{} 
     with two main steps: Data Collection from Website and Data Preprocessing. 
    }
    \label{fig:dataset}
    % \vspace{-0.5em}
\end{figure*}

\subsection{Task 3: Covalent Docking Pose Prediction}
\label{sec:Task 3}
The  final task is to predict the covalent docking pose, which is similar to traditional molecular docking tasks but includes the added complexity of a covalent bond.  Since the pocket and reactive site have been predicted in Task 1, here we formulate the covalent docking pose prediction (cov-docking) task as site-specific docking task.

Given the predicted pocket center $\mathcal{P}^c$ from Task 1, we circle around the pocket center with a radius $20$\AA~as the pocket, denoted by $\mathcal{P}^*$ and its structure position $\mathcal{R^{P^*}}$. With the predicted post-reactive ligand $\mathcal{M}'$ from Task 2, given its random conformation $\mathcal{R^{M'}}$, the goal is to predict the final docking pose $\mathcal{R^{M^*}}$ for the post-reactive ligand through a deep learning model $M_d$:
\begin{equation}
    \mathcal{R^{M^*}} = M_d(\mathcal{R^{M'}}, \mathcal{R^{P^*}}).
\end{equation}

For this task, we again use Uni-Mol~\cite{zhou2023}, which has demonstrated excellent performance in general molecular docking tasks. We modify the model to incorporate covalent constraints by adding an auxiliary loss $\mathcal{L}_{cov}$, which enforces the predicted distance between ligand atoms and reactive site residues to be minimized:
\begin{equation}
    \mathcal{L}_{cov} = \texttt{ReLU}(d_{ij} - \mathbf{D}_{inter}),
    \label{eq: loss-cov}
\end{equation}
where $d_{ij}$ is the covalent bond distance between the ligand atom $i$ and the reactive amino acid residue atom $j$, 
and $\mathbf{D}_{inter}$ is the predicted distance map for the distances between ligand and pocket atoms (the green matrix of the output for Cov-docking model as shown in Figure~\ref{fig:models}). 
This loss ensures that the predicted covalent bond is the shortest possible distance between ligand and pocket atoms.
Ideally, if the predicted covalent bond length/distance $d_{ij}$ is the closest one, the loss is zero. Otherwise, we push the covalent bond to be the shortest distance through this loss
\footnote{Empirically, we have counted all the ground-truth distance map for all the data samples and found the covalent bond distance was the shortest one. We have proven our hypothesis using the Kolmogorov-Smirnov test in Appendix~\opt{arxivVer}{\ref{sec: appendix-min-distance-statics}}}\opt{kddCRVer}{\href{\arxivurl}{E}}. 
Same as Uni-Mol, the main loss $\mathcal{L}_{dist}$ is the error between the predicted distance map and the ground-truth distance map, where the distance map consists of the intra distance $\mathbf{D}_{intra}$ within the ligand and the inter distance map $\mathbf{D}_{inter}$ between ligand and pocket atoms. Therefore, the final loss for covalent docking prediction is 
\begin{equation}
    \mathcal{L}_d = \mathcal{L}_{dist} + \mathcal{L}_{cov}.    
\end{equation}

Additionally, we include a post-process step for the predicted docking pose. We have analyzed all the covalent bond lengths, with their mean value $\mu$ and variance $\sigma$. We then randomly sample one value $l'$ from the normal distribution of $\mathcal{N}$($\mu$, $\sigma$), and if the predicted covalent bond length exceeds $l'$, we will move the ligand atom involved in the covalent bond toward the reactive aa to achieve the target bond length $l'$. Note that this post-process is optional, but we find it helpful for final performance.

\section{CovDocker: Benchmark Building}

\subsection{Dataset Collection}\label{sec:data-collection}
To construct a comprehensive benchmark dataset for covalent docking, as illustrated in Figure \ref{fig:dataset}, we systematically collected, processed, and filtered data from two primary sources: CovPDB~\cite{gao2022} and CovBinderInPDB~\cite{guo2022}. 
This process initially resulted in 2,754 entries for Task 2 and Task 3. For Task 1, we further refined the dataset by excluding proteins with chains exceeding 1,024 amino acids  within a 10 Å threshold of the ligand. These entries were discarded as they posed computational challenges for most deep learning models, leaving a total of 2,717 entries.

\textbf{Data Collection from Website.} 
In the first phase of data collection, we retrieved the latest available data (up to December 2023) from CovPDB~\cite{gao2022} and CovBinderInPDB~\cite{guo2022}. Due to the lack of critical information in some downloaded entries, we supplemented this dataset by including  additional PDB files from the Protein Data Bank~\cite{rcsb-pdb} and covalent bond information from CovPDB’s website. 
We then merged these datasets to ensure consistency. The final dataset includes essential data points for each entry, including the PDB ID, docked structure (comprising protein and ligand components), pre-reactive SMILES, and detailed covalent bond information, including the bond type, location, and involved atoms for both the ligand and the covalently bonded amino acid.
For further details, please refer to Appendix~\opt{arxivVer}{\ref{sec: appendix-fetch-data}}\opt{kddCRVer}{\href{\arxivurl}{A.1}}.

\textbf{Data Preprocessing.}
The preprocessing stage involves separating each complex into its ligand and protein components
For the protein, we isolate the active pocket within a 10Å radius of the ligand atoms. 
For the ligand, given that CovPDB does not provide post-reactive ligands and PDB files offer limited bond information, we adopt a CovBinderInPDB-inspired approach~\cite{guo2022} to generate post-reactive molecules. This process utilizes the Chemical Component Dictionary~\cite{lig-expo} to align the stable and isolated forms of post-reactive ligands with the covalently bonded forms of post-reactive ligands in PDB files. Special handling is implemented for several edge cases. For more details on preprocessing, refer to Appendix~\opt{arxivVer}{\ref{sec: appendix-preprocess}}\opt{kddCRVer}{\href{\arxivurl}{A.2}}.

\subsection{Dataset Split}
\textbf{Motivation for a New Test Set.} Traditional covalent docking methods are typically  evaluated using Keseru's benchmark evaluation set~\cite{scarpino2018}, which consists of  207 hand-curated high-resolution Cys-bound complexes based on published quality criteria. However, this might not reflect performance in real-world applications, where data might not be of approximately high quality and may be diverse in target reactive amino acids. Due to the differences in resolution and solely Cys-target covalent docking, as well as the strong dependence of docking success on the quality of experimental data observed in~\cite{goullieux2023}, the complexes of Keseru's benchmark evaluation set may be easier to predict than the average complex. Therefore, we introduce a new test set that better represents the challenges faced in practical applications.

\noindent\textbf{Dataset Split.} 
To ensure a fair evaluation and prevent data leakage from previous methods, we adopt the time-based split strategy used in EquiBind~\cite{stark2022}. 
Among the 2,754 preprocessed complexes, 2,308 were discovered before 2020. 
Entries from before 2020 are used for training, while data collected after January 1, 2020, are randomly split into validation and test sets.
This results in 2,308 training samples, 223 validation samples, and 223 test samples for Task 2 and Task 3. 
For Task 1, after applying further filtering (to remove proteins with a sequence length exceeding 1,024), we obtain 2,277 training samples, 220 validation samples, and 220 test samples. 
Importantly, to ensure consistency, the division of the validation and test sets remains the same across all tasks.

\subsection{Formulating Evaluation Metric}
For each task, we define specific metrics to evaluate model performance:

1. \textbf{Reactive Location Prediction}:

\begin{itemize}
    \item \textbf{Reactive Site Prediction}: \textit{AA\_Accuracy} is used to evaluate the prediction of the correct type and location of the reactive site (residue). The metric is defined as $AA\_accuracy=\#(pred\_idx==target\_idx)/\# (samples)$, where $\#()$ represents the count function.
    \item  \textbf{Pocket Prediction}: For pocket center prediction, we adopted the widely-used \textit{DCC} (Distance between the predicted pocket center and the native binding site’s center) metric~\cite{pei2024fabind}. 
\end{itemize} 

2. \textbf{Covalent Reaction Prediction}: We employ \textit{top-K exact match accuracy}, a standard metric in chemical reaction prediction tasks~\cite{irwin2022}. A prediction is considered correct if the true SMILES is included in the top-K predictions, with string matches checked after canonicalizing the SMILES.

3. \textbf{Covalent Docking Pose Prediction}:
\begin{itemize}
    \item We use root-mean-square deviation (\textit{RMSD}) between predicted and true ligand atomic coordinates, a standard metric for evaluating docking accuracy in noncovalent tasks~\cite{stark2022}.
    \item To better evaluate the model's performance in covalent bond formation, we introduce a new metric, \textit{RMSD (IB)}, which measures the RMSD between the covalently bonded ligand atom and the bonded protein atom. This metric provides a biologically relevant evaluation of the covalent bond's precision~\cite{oyedele2023}. 
\end{itemize}

\begin{table}[b]
\centering
\caption{Comparison of dataset availability, entry number, reactive mechanism types, reactive aa types with previous covalent docking methods (Open-Sourced Code Only) }
\label{tab: motivation-table}
\resizebox{\columnwidth}{!}{
    \begin{tabular}{lccccc}\toprule
    &\textbf{Dataset avail.} &\textbf{\# Entry} &\textbf{\# Mechanism type} &\textbf{\# Reactive aa type} \\\midrule
    \textbf{CovalentDock} &N &76 &2 &2 \\
    \textbf{Autodock4 (cov)} &N &20 &N.A. &2 \\
    \textbf{CDOCKER (cov}) &N &207 &7 &1 \\
    \textbf{CovDocker} &Y &2717/2754 &22 &10 \\
    \bottomrule
    \end{tabular}
}
% \vspace{-1em}
\end{table}

\subsection{Comparison with Existing Method}\label{sec: comparison}

We compare the dataset characteristics of \method{} against traditional covalent docking methods, focusing on dataset availability, entry numer, reaction type number and target aa type number of \method{}. A summary of these comparisons is provided in Table~\ref{tab: motivation-table}. Our analysis is restricted to methods with publicly available source code, in line with deep learning standards that emphasize reproducibility and verifiability. A comprehensive review of current covalent docking tools is provided in Appendix Table~\opt{arxivVer}{\ref{tab:review-covalent-docking-tools}}\opt{kddCRVer}{\href{\arxivurl}{8}}.

\textbf{Dataset Availability and Entry Number.} While the aforementioned methods have made their source code available, they only provide the PDB IDs of the datasets used rather than the complete processed datasets. In contrast, our \method{} provides full implementation details, including the complete processed dataset, model source code, and trained weights, ensuring full reproducibility. 
As shown in Table~\ref{tab: motivation-table}, our dataset significantly exceeds the scale of those used by CovalentDock~\cite{ouyang2013} , AutoDock4 (cov)~\cite{bianco2016}, and CDOCKER~\cite{wu2022a}. Notably, these traditional methods primarily utilize their limited datasets for evaluation purposes only, whereas our \method{} provides comprehensive coverage, including both training and evaluation datasets.

\textbf{Reactive Mechanisms and Amino Acid Types.} The distribution of reactive mechanisms and amino acid types are presented in Appendix Figures~\opt{arxivVer}{\ref{fig:mechanism-dist-n2754}}\opt{kddCRVer}{\href{\arxivurl}{6}}-\opt{arxivVer}{\ref{fig:aa-dist-n2754}}\opt{kddCRVer}{\href{\arxivurl}{7}} for the complete dataset and Appendix Figures~\opt{arxivVer}{\ref{fig:mechanism-dist-n2717}}\opt{kddCRVer}{\href{\arxivurl}{8}}-\opt{arxivVer}{\ref{fig:aa-dist-n2717}}\opt{kddCRVer}{\href{\arxivurl}{9}} for the 2,717 examples in Task 1. 
Our analysis reveals a diverse distribution of reactive mechanism, including Michael addition, ring opening, and other mechanisms, demonstrating the comprehensive coverage of binding mechanisms in our benchmark. Furthermore, the distribution of target amino acid types significantly surpasses that of previous works.
For comparison, CovalentDock only contains 63 $\beta$-lactam ligands bound to serine and 13 Michael acceptors bound to cysteines. AutoDock4 (cov) does not provide reaction mechanism information (denoted as N.A. in Table~\ref{tab: motivation-table}), while CDOCKER's dataset is exclusively focused on seven types of reaction mechanisms that specifically target cysteine residues. In contrast, our \method{} encompasses a substantially broader range of both reactive mechanism types and reactive amino acid types.

\section{Evaluation Results}
\subsection{Reactive Location Prediction}

\begin{table}[]
    \centering
    \caption{Results for reactive location prediction task.}
    \label{tab:reactive-site-res}
    \resizebox{\columnwidth}{!}{
        \begin{tabular}{@{}lcccc@{}}
        \toprule
        \textbf{}                  &  & \multicolumn{3}{c}{\textbf{\%DCC Below $\uparrow$}} \\ \cmidrule(l){3-5} 
        \textbf{Methods}          & \textbf{AA accuracy $\uparrow$}     & 3Å   & 4Å            & 5Å            \\ \midrule
        \textsc{Fpocket}          & -    & 5.0\%$\pm$0.0\% & 13.6\%$\pm$0.0\% & 18.6\%$\pm$0.0\%          \\
        \textsc{P2Rank}           & -    & 25.9\%$\pm$0.0\% & 55.0\%$\pm$0.0\% & 67.7\%$\pm$0.0\% \\
        \textsc{Ours}             & 63.5\%$\pm$3.0\% & 42.0\%$\pm$1.4\% & 51.4\%$\pm$2.0\% & 59.2\%$\pm$2.5\%
        \\ \bottomrule
        \end{tabular}
    }
\end{table}

\begin{table}[]
    \centering
    \caption{Results for covalent reaction prediction task.}
    \label{tab:reaction-pred-res}
    \resizebox{\columnwidth}{!}{
        \begin{tabular}{@{}lcccc@{}}
        \toprule
                             & \multicolumn{4}{c}{\textbf{Top-K accuracy $\uparrow$}}        \\ \cmidrule(l){2-5} 
        \textbf{Methods}     & 1             & 3             & 5             & 10            \\ \midrule
        \textsc{T5Chem}     & 55.0\% ± 0.9\% &65.8\% ± 0.7\% &68.6\% ± 1.2\% &71.7\% ± 0.4\%          \\
        \textsc{ReactionT5} & 71.2\% ± 1.0\% &80.7\% ± 0.4\% &81.8\% ± 0.5\% &83.4\% ± 0.4\% \\
        \textsc{Chemformer} & 76.7\% ± 0.4\% &83.0\% ± 0.0\% &83.1\% ± 0.3\% &83.7\% ± 0.3\%          \\ \bottomrule
        \end{tabular}
    }
\end{table}

\begin{table*}
\centering
\small
\caption{%Result for 
Comparison on the cov-docking task. RMSD (IB): Root Mean square Deviation between covalent inter-bond pocket atom and ligand aotm; Ours: Uni-Mol with $\mathcal{L}_{cov}$ auxiliary loss only; Ours-p: Uni-Mol with $\mathcal{L}_{cov}$ auxiliary loss and postprocess for inter bond ligand atom.}
\label{tab:cov-docking-res}
\resizebox{2\columnwidth}{!}{
    \begin{tabular}{@{}lccccccccc@{}}
    \toprule
                             & \multicolumn{4}{c}{\textbf{\% RMSD Below $\uparrow$}} & \multicolumn{4}{c}{\textbf{\% RMSD (IB) Below $\uparrow$}} &           \\ \cmidrule(lr){2-9}
    \textbf{Methods} &
      2Å &
      3Å &
      4Å &
      \multicolumn{1}{c|}{5Å} &
      0.5Å &
      1Å &
      2Å &
      \multicolumn{1}{c|}{3Å} &
      \multirow{-2}{*}{\textbf{\begin{tabular}[c]{@{}c@{}}Average \\ Runtime (s)\end{tabular}}} \\ \midrule
    \multicolumn{10}{l}{\cellcolor[HTML]{EFEFEF}\textit{Traditional methods}}                                                            \\
    \textsc{AutoDock4} (cov) & ~~~5.2\%$\pm$0.3\%*    & ~~20.3\%$\pm$2.2\%*   & ~36.9\%$\pm$2.5\%*   & \multicolumn{1}{c|}{~50.7\%$\pm$3.2\%*}  & -       & -       & -       & \multicolumn{1}{c|}{-}      & 167.23$\pm$1.05    \\ 
    \textsc{Vina}            & ~~6.1\%$\pm$0.5\%    & 17.0\%$\pm$2.0\%   & 22.4\%$\pm$2.5\%   & \multicolumn{1}{c|}{35.4\%$\pm$3.2\%}  & 0.0\%$\pm$0.0\% & 0.0\%$\pm$0.0\% & 7.0\%$\pm$0.3\%      & \multicolumn{1}{c|}{26.5\%$\pm$3.6\%}   & 20.25$\pm$0.36     \\
    \textsc{Smina}           & ~~9.9\%$\pm$0.4\% & 28.1\%$\pm$0.9\% & 38.0\%$\pm$1.1\%   & \multicolumn{1}{c|}{50.2\%$\pm$0.8\%}  & 0.0\%$\pm$0.0\% & 0.6\%$\pm$0.3\% & 12.9\%$\pm$1.4\%    & \multicolumn{1}{c|}{39.2\%$\pm$1.1\%}   & 20.33$\pm$0.06     \\ \midrule
    \multicolumn{10}{l}{\cellcolor[HTML]{EFEFEF}\textit{Deep learning-based methods}}                                                                                \\
    \textsc{Uni-Mol}         & 35.0\%$\pm$2.5\% & 51.4\%$\pm$0.3\% & 64.7\%$\pm$2.2\%   & \multicolumn{1}{c|}{74.6\%$\pm$0.9\%}  & 59.9\%$\pm$3.1\% & 72.6\%$\pm$2.0\% & 78.8\%$\pm$1.6\%    & \multicolumn{1}{c|}{83.7\%$\pm$2.8\%}   & 0.74$\pm$0.10 \\
    \textsc{Ours}          & 37.2\%$\pm$1.2\% & 53.8\%$\pm$0.9\% & 66.1\%$\pm$2.3\%   & \multicolumn{1}{c|}{74.1\%$\pm$1.8\%}  & 60.1\%$\pm$2.0\% & 73.1\%$\pm$2.5\% & 78.3\%$\pm$2.2\%    & \multicolumn{1}{c|}{83.6\%$\pm$1.3\%}   & 0.74$\pm$0.04 \\
    \textsc{Ours}-p & 37.2\%$\pm$1.2\% & 53.8\%$\pm$0.9\% & 66.2\%$\pm$2.5\% &
      \multicolumn{1}{c|}{75.9\%$\pm$1.9\%} & 79.1\%$\pm$0.7\% & 93.1\%$\pm$0.7\% & 98.4\%$\pm$0.7\% &
      \multicolumn{1}{c|}{100.0\%$\pm$0.0\%} &
      0.88$\pm$0.10 \\ \bottomrule
    \end{tabular}
}
\end{table*}

\textbf{Baselines.} For the pocket and reactive site prediction, we conducted comparative evaluations against P2Rank~\cite{krivak2018} and Fpocket~\cite{fpocket}, both of which are capable of predicting the protein pockets. However, neither of these tools can predict the reactive site. For our approach, we employ a ligand encoder initialized from the Uni-Mol pretrained checkpoint, while the protein encoder is trained from scratch. As discussed previously, we use a residue-level protein encoder specifically designed for this task. Details of the baseline models and the parameter settings can be found in Appendix~\opt{arxivVer}{\ref{sec: appendix-baseline-reactive-site}}\opt{kddCRVer}{\href{\arxivurl}{C}}. Since Fpocket and P2Rank do not predict the reactive site, their results on AA accuracy are omitted.

\noindent\textbf{Results.} As shown in Table \ref{tab:reactive-site-res}, our model is effective at capturing both the pocket and reactive site. Specifically, our pocket and site prediction model performs significantly better than the baselines in terms of \%DCC within 3Å. However, it performs on par with or slightly inferior than P2Rank at coarser resolutions like 4Å and 5Å. This discrepancy may be attributed to the fact that ligand-informed prediction are particularly effective at capturing binding patterns, but they may not generalize as well at higher resolutions.

\subsection{Covalent Reaction Prediction}
\textbf{Baselines.} For covalent reaction prediction evaluation, we compare our model three advanced models: T5Chem~\cite{lu2022b}, ReactionT5~\cite{sagawa2023}, and Chemformer~\cite{irwin2022}. 
All models are trained with their default parameter settings and fine-tuned on our dataset.
During the generation stage, we use a beam size of 10, and return 10 sequences for each model. 

\noindent\textbf{Results.}  Table \ref{tab:reaction-pred-res} presents the results of our comparative analysis. Chemformer achieves comparable performance to ReactionT5 while significantly outperforming T5Chem. Additionally,  
our results are much lower compared to the typically reported performance on the USTPO test set for predicting reaction products. This contrast indicates that our dataset, which includes organic reactions involving amino acids, presents greater complexity compared to conventional datasets that focus on small molecule systems.

\subsection{Covalent Docking}
\textbf{Baselines.} We compare our model with widely-used traditional docking methods, including Vina~\cite{vina}, Smina~\cite{smina}, and AutoDock4 (cov)~\cite{bianco2016}.  AutoDock4 (cov) is a covalent docking method, while Vina and Smina are commonly employed non-covalent methods. A brief review of the reproducibility of traditional covalent docking tools can be found  in Appendix~\opt{arxivVer}{\ref{sec: appendix-short-review-for-covalent-docking-tools}}\opt{kddCRVer}{\href{\arxivurl}{D}}. The ligand and pocket preparation process for non-covalent docking methods can be found in Appendix~\opt{arxivVer}{\ref{sec: appendix-ligand-prep-for-noncov}}\opt{kddCRVer}{\href{\arxivurl}{B.3}}. 
We use the default parameters for each tool, with a box length of 20 and an exhaustiveness of 8. For our deep learning baseline, we finetune Uni-Mol with training hyperparameters unchanged. 
Since AutoDock4 (cov) uses a flexible side-chain modeling method, the RMSD (IB) metric is not suitable in this case. Consequently, results of AutoDock4 (cov) in the RMSD (IB) column are omitted. 
The "*" indicates that due to the loss of bond information and index information in the docking results of AutoDock4 (cov), we cannot directly calculate the corresponding ligand RMSD. Instead, we use the Hungarian algorithm~\cite{kuhn1955hungarian} to reorder the docking results, which may result in slightly higher values than the actual ones.
The corresponding ligand RMSD values after reordering can be found in Appendix~\opt{arxivVer}{\ref{sec: appendix-reorder-res}}\opt{kddCRVer}{\href{\arxivurl}{F}}.

\noindent\textbf{Results.} The results, presented in Table \ref{tab:cov-docking-res}, can be summarized as follows:
(1) Overall, deep learning-based methods significantly outperform traditional methods in covalent docking.
(2) Our proposed $\mathcal{L}_{cov}$ auxiliary loss demonstrates effectiveness in improving both the RMSD and RMSD (IB) metrics. This suggests that our method substantially aids in transitioning a non-covalent approach into a covalent method. 
(3) The postprocessing step (as detailed in Section~\ref{sec:Task 3}) further enhances both RMSD and RMSD (IB) metrics (Our-p). 
(4) The traditional covalent docking method AutoDock4 (cov) performs poorly on our dataset, which may indicate that the given reactive residue alignment and flexible residue modeling method fail to fully capture the pocket information, leading to worse performance compared to upgraded non-covalent docking methods such as Vina and Smina.

\subsection{Pipeline Evaluation}

\begin{table}[b]
\centering
% \vspace{-1em}
\caption{Pipeline results for various covalent docking settings.}
\label{tab:res-pipeline}
\resizebox{\columnwidth}{!}{
    \begin{tabular}{@{}lllll@{}}
    \toprule
     &
      \multicolumn{2}{l}{\textbf{\begin{tabular}[c]{@{}l@{}}\%RMSD \\ Below $\uparrow$\end{tabular}}} &
      \multicolumn{2}{l}{\textbf{\begin{tabular}[c]{@{}l@{}}\% RMSD (IB) \\ Below $\uparrow$\end{tabular}}} \\ \cmidrule(l){2-5} 
    \textbf{Setting}              & 3Å   & \multicolumn{1}{l|}{5Å}   & 1Å   & 2Å   \\ \midrule
    \textsc{Site-specific} & 41.3\%$\pm$0.8\% & \multicolumn{1}{l|}{58.2\%$\pm$1.8\%} & 71.4\%$\pm$0.6\% & 75.4\%$\pm$0.9\% \\
    \textsc{Blind}         & ~~0.4\%$\pm$0.2\% & \multicolumn{1}{l|}{2.3\%$\pm$0.6\%} & 44.4\%$\pm$1.3\% & 47.0\%$\pm$1.7\% \\ \bottomrule
    \end{tabular}
}
\end{table}

\begin{figure*}
    \centering
    \includegraphics[width=0.8\textwidth]{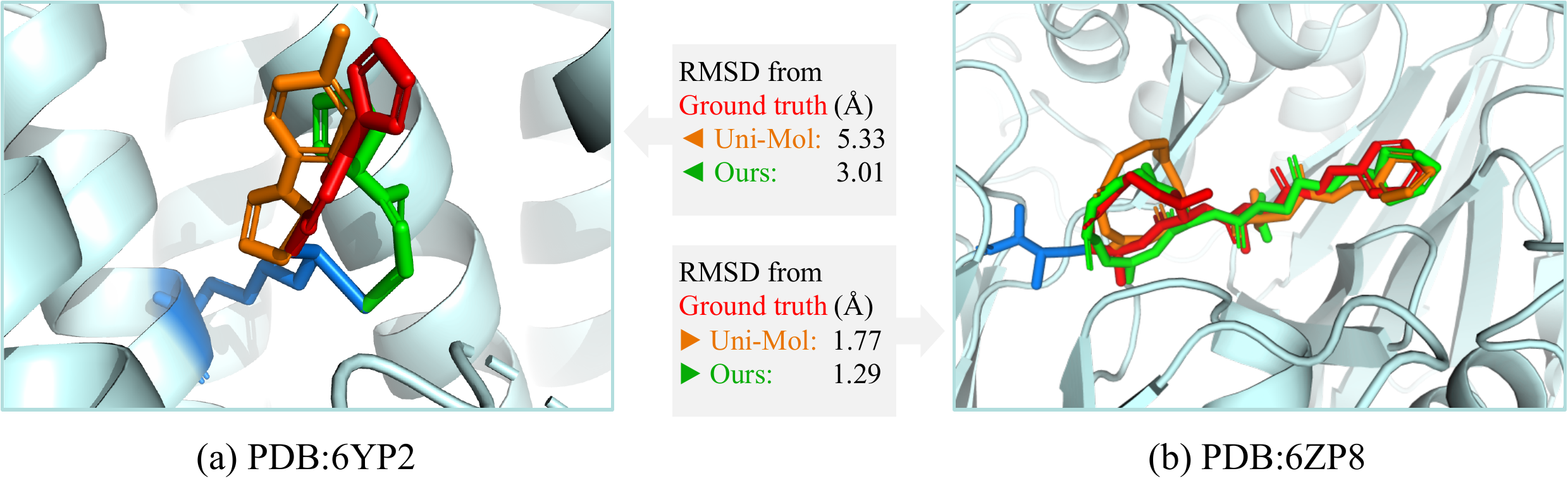}
    \caption{Visualization result of two case study. Structures predicted by Uni-Mol (orange) and ours (green) are placed together with the pocket target, with the RMSD to the ground truth (red) reported.}
    \label{fig:case}
\end{figure*}

While site-specific docking has proven effective in predicting ligand interactions when the target binding site is known, it pose limitations in blind docking scenarios where the binding site remains unidentified. 
To better assess the performance of our pipeline methods in a real-world, blind docking scenario, we conducted the covalent docking task using the inferred results of reactive site prediction and covalent reaction information obtained in earlier stages of the pipeline. 

The pipeline evaluation metrics for final docking performance are defined separately for blind and site-specific docking scenarios. 
For \textit{blind docking}, the score is calculated as the product of three components: reactive site prediction accuracy (AA accuracy from Task 1), covalent reaction prediction accuracy (Top-1 accuracy from Task 2), and the percentage of RMSD values below $N$\AA~ using the pocket center predicted in Task 1 (Task 3 results conditioned on Task 1 predictions). 
For \textit{site-specific docking}, the score consists of two components: the covalent reaction prediction accuracy (Top-1 accuracy from Task 2) and the percentage of RMSD values below $N$\AA~ (Task 3 results).
As shown in Table \ref{tab:res-pipeline}, although our covalent docking model achieves good performance based on the covalent reaction prediction task (namely site-specific covalent docking), the blind docking results are suboptimal due to cumulative errors in the pipeline.

\subsection{Ablation Study}

We conducted ablation studies to evaluate the impact of key components in Task 1 and 3. No ablation study was performed for Task 2, as it involved only three existing baseline methods.

\textbf{Task 1}
Table~\ref{tab:reactive-site-ablation} analyzes two key components: (1) \textit{EmbFuse}, which derives the fused embedding from protein and ligand embeddings, and (2) \textit{Pocket Center}, which predicts pocket center coordinates from the fused embedding. (a) The first row shows the method used in our main text. (b) \textit{Concat Pooling} fuses embeddings by repeating, concatenating, and pooling protein and ligand embeddings. (c) \textit{CLS Coord} predicts 3D coordinates directly from the fused CLS embedding, while \textit{Weighted Coord} uses the fused embedding to weight protein coordinates for prediction. Results show that \textit{Weighted Coord} outperforms existing methods, and \textit{Cross Attention} effectively integrates protein and ligand embeddings, enhancing performance.

\textbf{Task 3} We evaluated four methods, summarized in Table~\ref{tab:cov-dock-ablation}: 
(a) “Ours-p” (main text method), (b) “Ours” (main text method), 
(c) replacing $L_{cov}=\text{ReLU}(d_{ij}-D_{inter})$ with $L_{cov(baseline)}=\Vert d_{ij}-min(D_{inter})\Vert$, and  
(d) no $L_{cov}$ and post-processing (equivalent to Uni-Mol). The ablation results show that $L_{cov}$  significantly improves RMSD, while post-processing 
plays a critical role in enhancing RMSD (IB) performance.

\begin{table}[]
    \centering
    \caption{Ablation study on task 1.}
    % \vspace{-1em}
    \label{tab:reactive-site-ablation}
    \resizebox{\columnwidth}{!}{
        \begin{tabular}{llllll}
        \toprule
            & EmbFuse         & Pocket center  & AA Accuracy & \%DCC\textless{}3 Å& \%DCC\textless{}4 Å\\ \midrule
        (a) & Cross Attention & Weighted Coord & 65.0\%      & 43.2\%            & 52.3\%            \\
        (b) & Concat Pooling  & Weighted Coord & 52.3\%      & 43.2\%            & 50.9\%            \\
        (c) & Cross Attention & CLS Coord      & 23.6\%      & ~~0.9\%             & ~~1.8\%    
        \\ \bottomrule
        \end{tabular}
    }
    % \vspace{-1em}
\end{table}

\begin{table}[]
    \centering
    \caption{Ablation study on task 3.}
    % \vspace{-1em}
    \label{tab:cov-dock-ablation}
    \resizebox{\columnwidth}{!}{
        \begin{tabular}{@{}lcccc@{}}
        \toprule
                   & $L_{cov}$           & Postprocess & \%RMSD\textless{}2.0 Å & \% RMSD (IB) \textless 0.5 Å            \\ \midrule
                (a) & Yes                  & Yes         & 37.7\%                 & 77.1\%                       \\
        (b) & Yes                  & No          & 37.7\%                 & 59.2\%                       \\
        (c) & $L_{cov(baseline)}$ & No          & 34.5\%                 & 61.0\%                       \\
        (d) & No                   & No          & 32.3\%                 & 57.0\%               \\ \bottomrule
        \end{tabular}
    }
    % \vspace{-0.5em}
\end{table}

\subsection{Visualization}

We present visualizations for site-specific covalent docking in Figure \ref{fig:case}. The covalent bond is depicted in blue, with both our method and the ground truth explicitly connecting to the pocket via this covalent bond. In contrast, Uni-Mol, a non-covalent docking method, lacks this explicit connection. In both cases, our method outperforms Uni-Mol in terms of the RMSD metric. Notably, in the case of 6YP2, the Uni-Mol model (orange) predicted the entire molecule's orientation incorrectly compared to ground truth (red), which leads to a wrong pose. However, with the inclusion of our auxiliary loss, which enforces geometric constraints during covalent bond formation in Ours (green), the orientation was corrected, yielding a better predicted pose. Similarly, for 6ZP8, the auxiliary loss helped achieve a smaller RMSD in the docking results.

\section{Conclusion}
Covalent docking poses unique challenges due to the intricate formation of covalent bonds between molecules, which involve not only the prediction of binding sites but also structural transformations. In this study, we have established a comprehensive benchmark tailored for deep learning approaches, providing carefully crafted tasks and high-quality datasets to improve the prediction of binding behaviors in covalent drug design. By integrating advanced models such as Uni-Mol and Chemformer, we have enhanced the accuracy of reactive site prediction and model structural transformations that occur during covalent bonding, thus advancing the state of the art in this area. 
Our work lays a robust foundation for future research in covalent docking and drug discovery, enabling the development of more selective and potent covalent inhibitors. The benchmark we propose offers a structured and reproducible framework, which will serve as a valuable tool for researchers working to tackle the complexities of covalent docking. Moreover, it underscores the potential of tailored computational approaches to drive innovation in chemical biology and drug design.

\begin{acks}
This work is supported by Key-Area Research and Development Program of Hubei Province (2024BCB027).
\end{acks}

\bibliographystyle{ACM-Reference-Format}
\balance
\bibliography{ref}

% \citestyle{acmauthoryear} % uncomment this if you need this type of citation

\opt{arxivVer}{
\newpage
\appendix
\nobalance
\section*{Appendix}
 % This appendix provides detailed supplemental material on dataset construction, preprocessing, and modeling for covalent docking. It covers data retrieval, molecule preparation, and complex PDB file processing. Additionally, it outlines methods for predicting docking poses and reactive sites, discusses baselines for traditional covalent docking tools, and addresses limitations encountered in our analyses.
\section{Dataset Construction}
\subsection{%Details on 
Fetching Data from Website}\label{sec: appendix-fetch-data}
\paragraph{Details on CovPDB and CovBinderInPDB datasets.} CovPDB contains PDB data collected up to August 31, 2020, encompassing 733 proteins, 1,501 ligands, and 2,294 covalent protein–ligand complexes. The dataset includes 14 targetable residues and 21 covalent reaction mechanisms. In contrast, CovBinderInPDB, which incorporates PDB data up to January 18, 2022, comprises 1,170 protein chains, 2,189 ligands, and 3,555 complex structures, along with 9 targetable residues and 113 reaction patterns. The differences in dataset size and composition reflect variations in their respective data collection strategies.

\paragraph{Data Download and Web Crawling.}We obtained CovPDB \footnote{\url{https://drug-discovery.vm.uni-freiburg.de/covpdb/download}} and CovBinderInPDB\footnote{\url{https://yzhang.hpc.nyu.edu/CovBinderInPDB/}} dataset from their official website. For CovPDB, since the HET codes of the bound ligands in the PDB files were not provided, we extracted this information by crawling the dataset's complex list available at \url{https://drug-discovery.vm.uni-freiburg.de/covpdb/complexes_list/initial=Allsortedby=protein_id}. For CovBinderInPDB, since the PDB structure files were not provided, we used the PDB fetch tool~\cite{pdbfetch} to download the corresponding PDB files. We then merged the data from CovPDB and CovBinderInPDB. In cases where entries appeared in both datasets, we prioritized the data from CovPDB due to its higher quality~\cite{goullieux2023}. This merging process resulted in a consolidated dataset of 4,064 examples. To identify covalent bonds, we extracted information from the \verb|LINK| records in the PDB files. Finally, we retained only those structures that contained exactly one protein-ligand covalent bond.

\paragraph{Gain Pre-reactive Molecule.} For CovBinderInPDB, the authors provided the pre-reactive molecule's SMILES expression in downloaded excel file from their official website. For CovPDB, the authors provided the pre-reactive molecule's SDF format file. So we additionally used \verb|rdkit| to get the SMILES expression from SDF file. 
% We filtered out ligands containing either Palladium (Pd) or Gold (Au), resulting in the exclusion of two ligands.

\subsection{%Details on 
Pre-processing Details}\label{sec: appendix-preprocess}
\paragraph{Splitting Complex PDB Files into Protein and Ligand Components.}  We utilized the bounded ligand's HET code to identify the target ligand of interest, filtering out all other clashes and water molecules in the PDB file. Furthermore, we exclude entries that have more than one linked residue to the target ligand. We retained only those complex structures whose ligand files, in SDF or MOL2 format, could be processed by RDKit and OpenBabel. 

\paragraph{Select Protein Chains and Pocket Atoms.} Our pocket radius is set to 10 Å. To prevent the input residue number for the reactive location prediction model from being too long, we filter out chains where the minimum contact distance between amino acid atoms and ligand atoms exceeds the 10 Å cutoff. We retained only those proteins that have an atom within a 10 Å radius of any ligand atom. We also filtered out complexes with alternate positions for ligands or bonded amino acids, retaining only one binding scenario for symmetric ligands and receptors. Furthermore, we excluded complexes that still contain more than 1,022 residues after the aforementioned filtering process. We set 1,022 as the maximum length to reserve two token positions for beginning-of-sequence (BOS) and end-of-sequence (EOS) tokens.

\paragraph{Gain Post-reactive Molecule.} The \verb|CONNECT| field specifies connectivity but does not distinguish between single and double bonds. Moreover, since hydrogen atoms are not provided in the PDB files, it is impossible to deduce the corresponding connections through the valency of the atoms. Thus, we adopted a workflow similar to that of CovBinderInPDB to derive the post-reactive molecules. First, We constructed SDF files based on the mmCIF files of the stable form of the bound ligand when isolated, as provided in the Chemical Component Dictionary\footnote{Here we used Ligand Expo\cite{lig-expo}.}. We then compared these with the structure of the post-reactive ligand in PDB file, removing extra atoms from the constructed SDF files. This includes removing extra non-hydrogen atoms in the Chemical Component Dictionary and one hydrogen atom connected to the bonded ligand atom. Subsequently, we connected corresponding parts of the amino acids to these structures. We then applied patches for several special cases: converting double/triple bonds to single/double bonds and adding a hydrogen atom. Additionally, we marked the boron in the boronation reactions with a negative charge. This post-reactive molecule process is applied to the data from both CovPDB and CovBinderInPDB to ensure consistency. We retained only those ligands that could be aligned between the Chemical Component Dictionary and the PDB file.

\balance

\subsection{Statistics on the Crafted Datasets}
The statistical results of amino acid types in our dataset are illustrated in Figure~\ref{fig:mechanism-dist-n2754} and Figure~\ref{fig:mechanism-dist-n2717}. Our analysis reveals a diverse distribution of reactive mechanisms, including Michael addition, ring opening, and others, highlighting the comprehensive coverage of binding mechanisms in our benchmark. Moreover, the distribution of target amino acid types significantly exceeds that of previous works. As shown in  Figure~\ref{fig:aa-dist-n2754} and Figure~\ref{fig:aa-dist-n2717}, the reactive amino acid types are far more extensive compared to earlier studies primarily focused on Cys or Ser~\cite{scarpino2018, wen2019}.

\section{%Details of the 
Modeling for Each Task}
\label{sec:app_models}

All our experiments were run on 8 NVIDIA V100 16GB GPUs.

\subsection{Reactive location Prediction}

\paragraph{Model Details.} Our implemented reactive site model can be viewed in Figure \ref{fig:models}. The ligand encoder is finetuned from Uni-Mol pretrained checkpoint on 209M molecular conformations. The protein encoder is randomly initialized since our protein encoder is a residue-level model, whereas Uni-Mol's pocket model is an atom-level model. In order to merge features from both protein and ligand, we utilize one cross-attention layer, where the query is from the protein, and the key and value are from the ligand, to get the output of the same length as the protein. We refer to the cross-attention implementation of Stable Diffusion in~\cite{diffusers}. The output complex embedding from the cross-attention layer is passed through a sigmoid layer to get the weight value of the current amino acid. The final pocket center is a weighted average of all input amino acids' C$_\alpha$ coordinates as 
\(
    \mathcal{P}^c = \frac{\sum_j p_j\times x_j}{\sum_j p_j}
\)
where \(p_j\) is the predicted pocket probability for \(j\)-th amino acid.

\begin{figure}[H]
    \centering
    \includegraphics[width=\columnwidth]{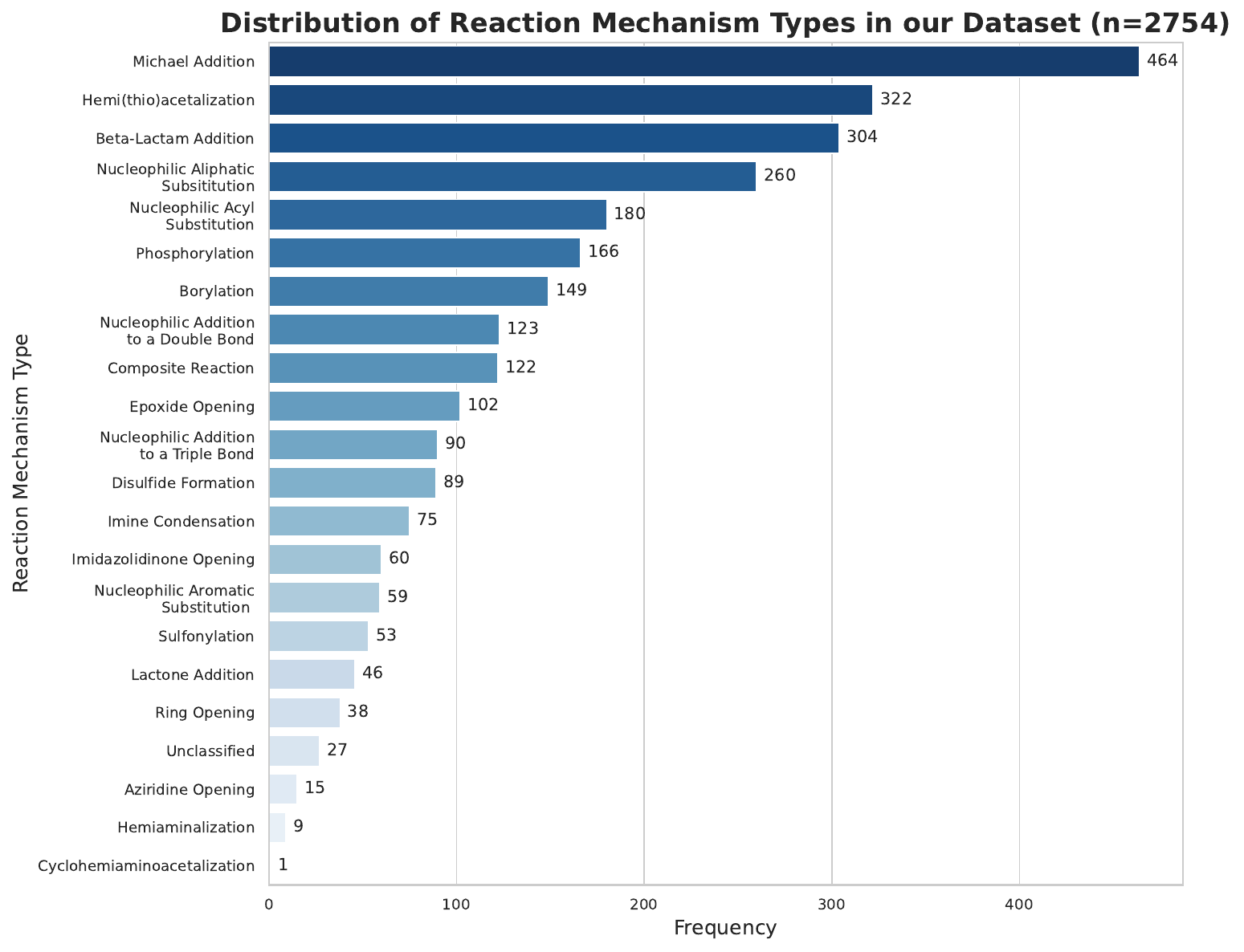}
    \caption{Reactive Mechanism Type Statistics (n=2754)}
    \label{fig:mechanism-dist-n2754}
\end{figure}

\begin{figure}[H]
    \centering
    \includegraphics[width=\columnwidth]{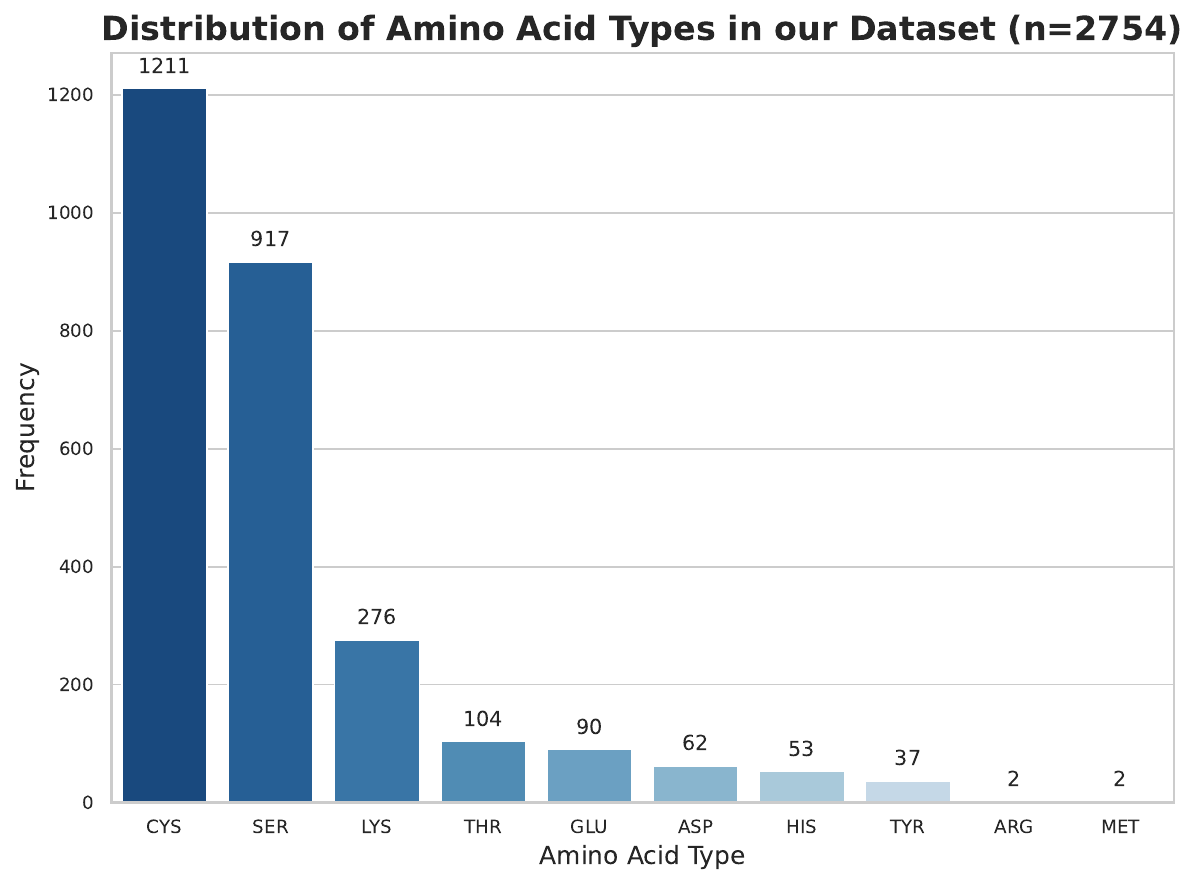}
    \caption{Reactive Amino Acid Type Statistics (n=2754)}
    \label{fig:aa-dist-n2754}
\end{figure}

\begin{figure}[H]
    \centering
    \includegraphics[width=\columnwidth]{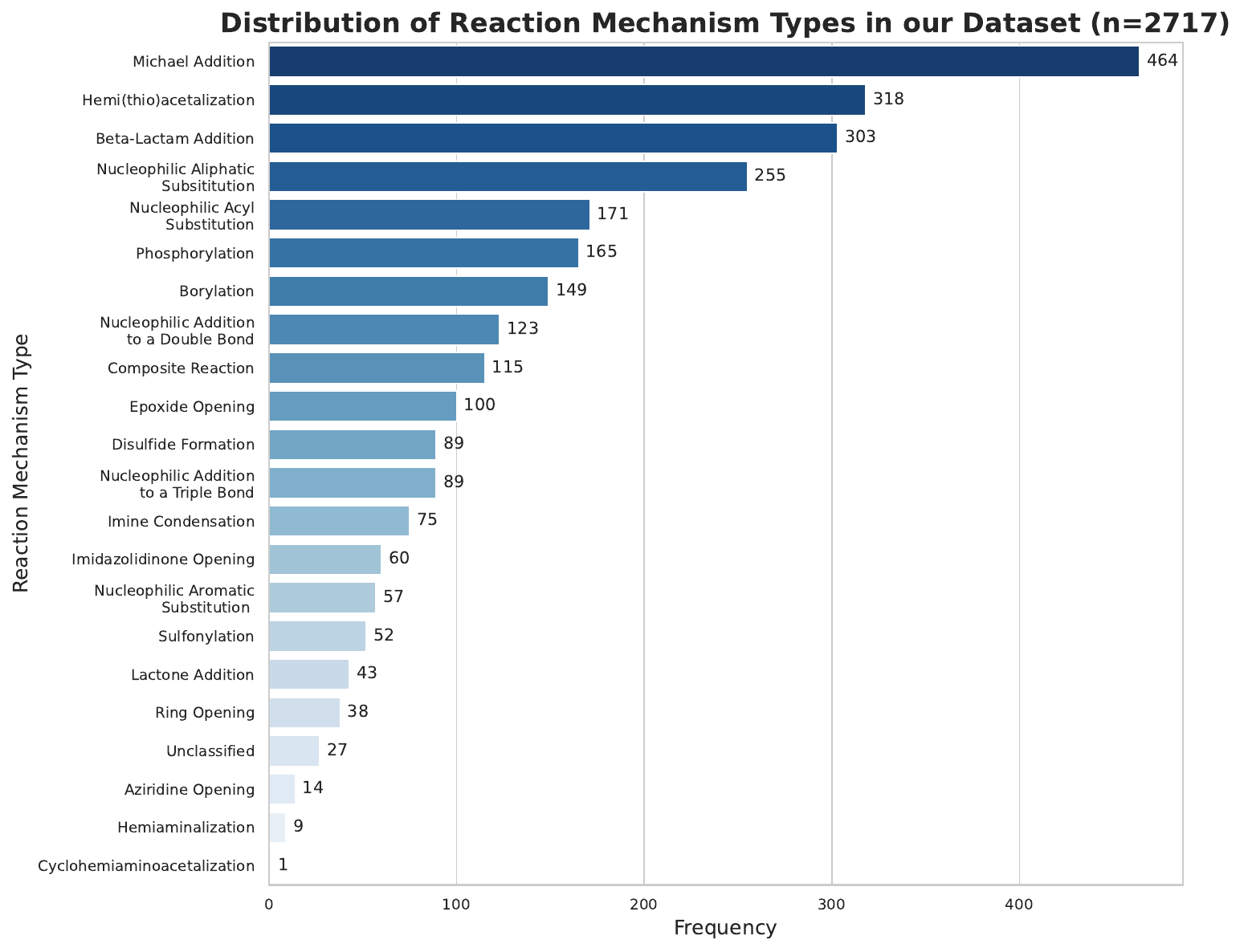}
    \caption{Reactive Mechanism Type Statistics (n=2717)}
    \label{fig:mechanism-dist-n2717}
\end{figure}

\begin{figure}[H]
    \centering
    \includegraphics[width=\columnwidth]{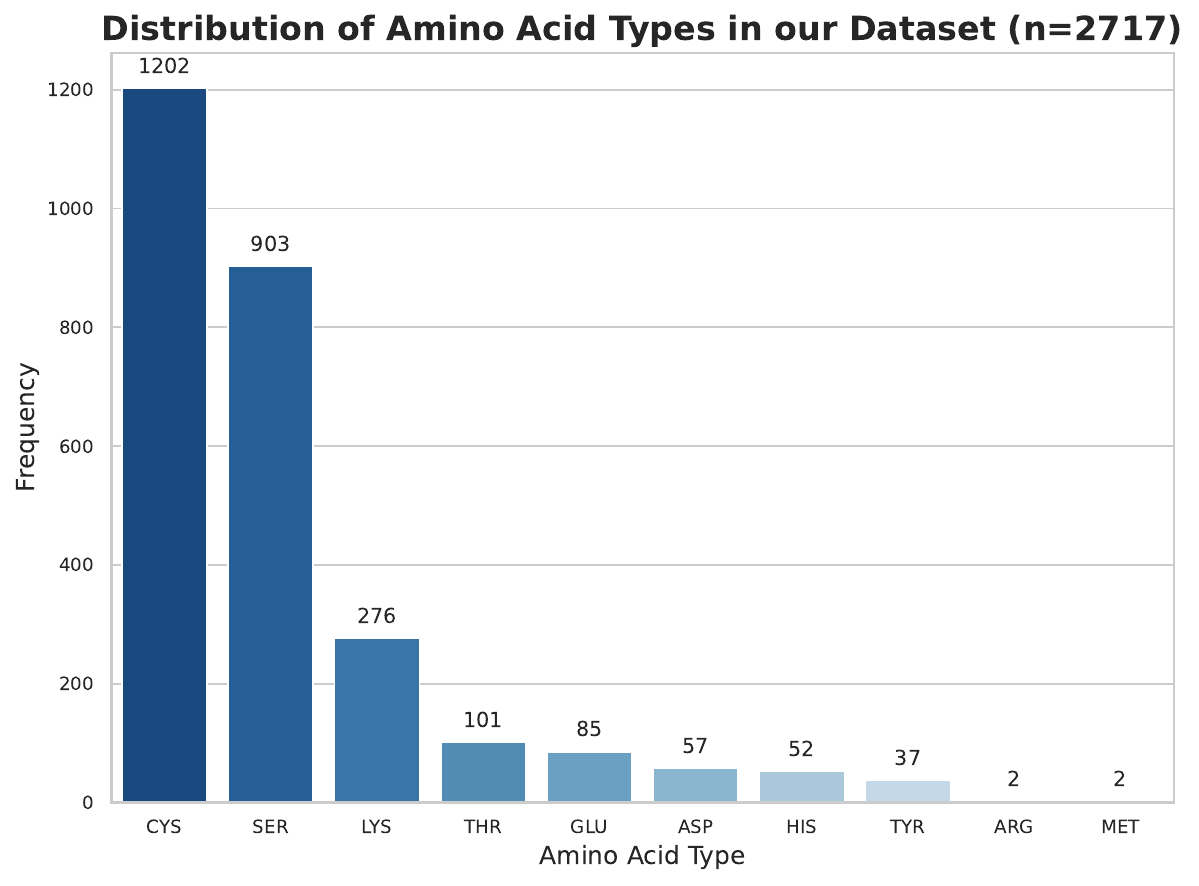}
    \caption{Reactive Amino Acid Type Statistics (n=2717)}
    \label{fig:aa-dist-n2717}
\end{figure}

\paragraph{Hyperparameters.} Huber loss~\cite{huberloss} is used as the loss function for the pocket prediction subtask. During the inference stage, the pocket residues are identified by the 20Å distance threshold from the predicted pocket center. For the reactive amino acid prediction subtask, the model output is a sequence-level classification result based on the pocket residues' complex embedding from the cross-attention layer's output, and cross-entropy loss is used as the loss function. Instead of directly using the model-predicted pocket residues as input for the reactive amino acid prediction subtask during training, which disobeys the teacher forcing style training~\cite{teacher-forcing}, we use the ground truth pocket residues during training and the predicted pocket residues during inference. Adam~\cite{adam} is used as the optimizer when training our model and betas in Adam are 0.9 and 0.99 for all experiments. The protein model consists of 25 layers of Uni-Mol blocks. Loss weight \(\alpha\) for reactive amino acid prediction subtask is 0.05 for our model.
% We further implement two sizes of the protein encoder model. The base model consists of 15 layers of Uni-Mol blocks, consistent with the Uni-Mol pretrained molecular model. 

\begin{table}[]
\centering
\caption{An overview on current covalent docking tools.}
\label{tab:review-covalent-docking-tools}
\resizebox{0.9\columnwidth}{!}{
\begin{tabular}{lccc}
\hline
\textbf{Method} & \textbf{Publish year} & \textbf{Open-sourced code} & \textbf{Public accessible} \\ \hline
\rowcolor[HTML]{D9D9D9} 
GOLD (cov)\cite{verdonk2003}           & 2003 &  \ding{55} & \ding{55} \\
ICM-PRO~\cite{katritch2007}        & 2007 & \ding{55} & \ding{55} \\
\rowcolor[HTML]{D9D9D9} 
FITTED (cov)~\cite{moitessier2016}        & 2012 & \ding{55} & \ding{55} \\
CovalentDock~\cite{ouyang2013}   & 2013 & \ding{51} & \ding{51} \\
\rowcolor[HTML]{D9D9D9} 
CovDock ~\cite{zhu2014}       & 2014 & \ding{55} & \ding{55} \\
DOCKovalent ~\cite{london2014}   & 2014 & \ding{55} & \ding{55}* \\
\rowcolor[HTML]{D9D9D9} 
DOCKTITE~\cite{scholz2015}       & 2015 & \ding{51}* & \ding{55} \\
AutoDock4 (cov)~\cite{bianco2016} & 2016 & \ding{51} & \ding{51} \\
\rowcolor[HTML]{D9D9D9} 
MOE (cov)\cite{moe}             & 2018 & \ding{55} & \ding{55} \\
WIDOCK  ~\cite{scarpino2021}       & 2021 & \ding{55} & \ding{55} \\
\rowcolor[HTML]{D9D9D9} 
COV\_DOX~\cite{wei2022}       & 2022 & \ding{55} & \ding{55} \\
CDOCKER (cov) ~\cite{wu2022a}       & 2022 & \ding{51} & \ding{51} \\
\rowcolor[HTML]{D9D9D9} 
HCovDock ~\cite{wu2023}      & 2023 & \ding{55} & \ding{51} \\
AC (cov) ~\cite{goullieux2023}      & 2023 & \ding{55} & \ding{51} \\ \hline
\end{tabular}}
\end{table}

\subsection{Covalent Reaction Prediction}
% detail modeling and the loss formulation
Developed in 2022, T5Chem utilizes the Text-to-Text Transfer Transformer (T5) framework~\cite{T5} and self-supervised pretraining with PubChem molecules for diverse chemical reaction predictions, including reaction type classification and synthesis predictions. ReactionT5, introduced in 2023, extends this model by pretraining on the Open Reaction Database (ORD) to improve yield and product predictions. Chemformer also leverages a Transformer-based architecture and self-supervised pretraining to achieve top-tier accuracy in synthesis tasks. In our implementation, we used the default hyperparameters and model settings for each model.

\subsection{Covalent Docking Pose Prediction}
\paragraph{Ligand and Pocket Preparation for Noncovalent Docking.}\label{sec: appendix-ligand-prep-for-noncov} To obtain the input ligand for noncovalent models from the output of the reaction model, which contains both the bonded amino acid part and the bonded ligand part, we further process the post-reactive ligand SMILES as follows. First, we remove the amino acid parts from the post-reactive ligand's SMILES. Next, we align this SMILES with the corresponding post-reactive ligand's PDB file to synchronize the bond order and coordinate information. Finally, we employed ETKDG~\cite{ETKDG} algorithm using RDKit~\cite{RDKit} to randomly generate a low-energy conformation as the initial pose of the ligand. To obtain the input pocket for docking models, we used OpenBabel~\cite{openbabel} to obtain coordinates for all pocket atoms, following the same criterion for pocket atom identification as in reactive site prediction task.

\paragraph{Loss Function.} We employed \(\mathcal{L}_{dist}\) for the same as Uni-Mol to use \(\mathcal{L}_{dist}\)=SmoothL1Loss \((D_{intra}, D_{intra}')\) + MSELoss \((D_{inter}, D_{inter}')\) as the main loss, where \(D_{intra}\) is the distance map between ligand atoms and ligand atoms, while \(D_{inter}\) is the distance map between ligand atoms and pocket atoms. The prime version denotes the target values. The auxiliary \(\mathcal{L}_{cov}\) loss is demonstrated in Equation \ref{eq: loss-cov}.

\paragraph{Postprocessing Details.} If the covalent inter-bond length between the bonded ligand atom and the bonded pocket atom deviates more than 10$\sigma$ from the mean covalent inter-bond length (where $\sigma$ is the standard deviation), the bonded ligand atom will be moved to a new coordinate to ensure the covalent inter-bond length satisfy the normal distribution.

\paragraph{Hyperparameters.} We used the default hyperparameters as Uni-Mol docking model. We experimented with adjusting the loss weight for \(\mathcal{L}_{cov}\), but initial experiments indicated that maintaining the loss weight at 1 yielded better results.

\section{Baselines for Reactive Location Prediction} \label{sec: appendix-baseline-reactive-site}
P2Rank, an open-source software available as both a standalone command line program and a Java library, features a user-friendly, lightweight installation and does not require other bioinformatics tools or extensive databases. It is noted for rapid processing and automated predictions, making it ideal for large datasets and scalable bioinformatics pipelines. Conversely, Fpocket employs a fast geometric method involving the filtering and clustering of alpha spheres via Voronoi tessellation. This tool is favored for its extensive use in large-scale projects and its ability to generate a high number of predicted pockets per protein.
We used the default parameter for each tool. To obtain the final pocket center coordinate, for P2Rank, we chose the rank 1 predicted pocket center result. For Fpocket, since it does not output a pocket center coordinate, we chose the rank 1 predicted atoms contacted by alpha spheres in the given pocket and take the geometric center of these atoms as the predicted pocket center to calculate the DCC metric.

\section{Baselines for Traditional Covalent Docking Tools} \label{sec: appendix-short-review-for-covalent-docking-tools}
A short review on current covalent docking tools can be seen in Table~\ref{tab:review-covalent-docking-tools}. The check in the "Publicly Accessible" column indicates that either their code is publicly accessible or they have published a website server. DOCKovalent's website is still accessible, but the docking service is already down. DOCKTITE is written as an SVL script for MOE. Even though DOCKTITE open-sourced their script code, MOE is not publicly accessible, so DOCKTITE is still not publicly accessible.

When compared to traditional covalent docking tools, we excluded closed-source and web server-based tools due to their limited accessibility and lack of transparency, which are essential for reproducible benchmarks. This decision ensures our benchmark remains robust, accessible, and adaptable for in-depth evaluation. Due to significant challenges with the runtime environments of CovalentDock and CDOCKER, as well as the fact that CovalentDock is outdated and no longer maintained, we were unable to successfully run these tools. As a result, AutoDock4 (cov) was the only traditional covalent docking tool used as a baseline in our experiment.

\begin{table*}[h]
\centering
% \small
% \vspace{-0.5cm}
\caption{Results for cov-docking task after reordering}
\label{tab:cov-docking-res-reorder}
\resizebox{1.8\columnwidth}{!}{
    \begin{tabular}{@{}lcccccccc@{}}
        \toprule
                                 & \multicolumn{4}{c}{\textbf{\% RMSD Below $\uparrow$}} & \multicolumn{4}{c}{\textbf{\% RMSD (IB) Below $\uparrow$}} \\ \cmidrule(lr){2-9}
        \textbf{Methods} &
          2Å &
          3Å &
          4Å &
          \multicolumn{1}{c|}{5Å} &
          0.5Å &
          1Å &
          2Å &
          \multicolumn{1}{c}{3Å} \\ \midrule
        \multicolumn{9}{l}{\cellcolor[HTML]{EFEFEF}\textit{Traditional methods}}                                                            \\
        \textsc{AutoDock4} (cov) & ~~~~5.2\% ± 0.3\% &20.3\% ± 2.2\% &36.9\% ± 2.5\%   & \multicolumn{1}{c|}{~50.7\% ± 3.2\%}  & -       & -       & -       & \multicolumn{1}{c}{-}      \\ 
        \textsc{Vina}            & ~~~~7.9\% ± 0.9\% &23.5\% ± 1.4\% &40.4\% ± 0.8\%   & \multicolumn{1}{c|}{55.3\% ± 1.1\%}  & 0.0\% ± 0.0\% &0.0\% ± 0.0\% &7.0\% ± 0.3\%      & \multicolumn{1}{c}{26.5\% ± 3.6\%}   \\
        \textsc{Smina}           & ~13.9\% ± 0.8\% &35.0\% ± 1.6\% &57.1\% ± 0.7\%    & \multicolumn{1}{c|}{75.8\% ± 1.2\%}  & 0.0\% ± 0.0\% &0.6\% ± 0.3\% &12.9\% ± 1.4\%    & \multicolumn{1}{c}{39.2\% ± 1.1\%}   \\ \midrule
        \multicolumn{9}{l}{\cellcolor[HTML]{EFEFEF}\textit{Deep learning-based methods}}                                                                                \\
        \textsc{Uni-Mol}         & 40.1\% ± 4.4\% &59.5\% ± 1.0\% &79.2\% ± 0.7\%   & \multicolumn{1}{c|}{89.5\% ± 1.1\%}  & 59.9\% ± 3.1\% &72.6\% ± 2.0\% &78.8\% ± 1.6\%    & \multicolumn{1}{c}{83.7\% ± 2.8\%}   \\
        \textsc{Ours}          & 42.0\% ± 0.7\% &60.7\% ± 1.4\% &79.2\% ± 3.2\%    & \multicolumn{1}{c|}{90.4\% ± 0.7\%}  & 60.1\% ± 2.0\% &73.1\% ± 2.5\% &78.3\% ± 2.2\%    & \multicolumn{1}{c}{83.6\% ± 1.3\%}   \\
        \textsc{Ours}-p & 42.0\% ± 0.7\% &61.0\% ± 1.2\% &79.8\% ± 3.9\% &
          \multicolumn{1}{c|}{90.6\% ± 0.4\%} & 79.1\% ± 0.7\% &93.1\% ± 0.7\% &98.4\% ± 0.7\% &
          \multicolumn{1}{c}{100.0\% ± 0.0\%} \\ \bottomrule
    \end{tabular}
}
\end{table*}

\paragraph{AutoDock4 (cov) Baseline.} Two different covalent docking procedures were implemented in AutoDock4 (cov)~\cite{bianco2016}. We chose the flexible side chain method as it performed better~\cite{bianco2016}. The use of this method requires the addition of two link atoms from the target residue to the ligand. So the input is further processed based on the input to non-covalent docking model to keep the additional two atoms from the post-reactive ligand. Close attention must be paid to the geometry of the initial modification, or the docking process may fail. The following steps are the same as their default processing workflow. During the docking, the protein and the backbone of its reactive site are treated as rigid and the C$\alpha$ and C$\beta$ atoms are fixed. The ligand is considered a fully flexible extension of the protein reactive site.

\section{Analysis on the Minimum Distance in Inter Distance Map}\label{sec: appendix-min-distance-statics}
To demonstrate the rationale of our covalent auxiliary loss $\mathcal{L}_{cov}$, we performed a statistical analysis on the relationship between the minimum distances in the inter distance map and the covalent inter-bond distances. 

To test whether the minimum distances in the inter distance map and covalent inter-bond distances are drawn from the same distribution, we performed a Kolmogorov-Smirnov test~\cite{Hodges1958TheSP} on 2754 data points. The resulting p-value is 0.9999999. The p-value is greatly above the usual confidence level threshold of 0.05, so we cannot reject the null hypothesis that the two samples were drawn from the same distribution. Furthermore, the frequency of covalent inter-bond distances being the minimum distance in the inter distance map is 0.996. This frequency observation also supports our hypothesis that covalent inter-bond distances should be the minimum distances in the inter distance map.

\section{Results of Covalent Docking after Reordering using Hungarian Algorithm} \label{sec: appendix-reorder-res}
Table \ref{tab:cov-docking-res-reorder} shows the ligand RMSD results after reordering using Hungarian algorithm. From the table we can see that AutoDock4 (cov) is actually the worst.

\section{Limitation}
\label{sec:app_limitation}
Covalent docking presents significant challenges compared to traditional non-covalent docking, particularly in problem modeling and algorithm design, leading to poor performance of current methods on such tasks. Our study has made meaningful progress in covalent docking methodologies. However, we keep only one binding scenario for symmetric ligands and receptors, which may introduce significant discrepancies compared to real data, particularly in blind covalent docking scenarios. Additionally, our approach does not adequately model internal covalent bonds within ligands or between pocket atoms. Our current setting uses a flexible ligand and a rigid pocket, which limits the real-world applicability of our results, as most real-world scenarios involve flexible pockets. Future research should focus on developing more integrated, end-to-end solutions to reduce these cumulative errors and enhance the flexibility of both ligands and pockets to better mimic real biochemical interactions.

}

\end{document}